\documentclass[aps,pra,twocolumn,showpacs,superscriptaddress ]{revtex4-1}
\usepackage{enumerate}
\usepackage{mathrsfs}
\usepackage{amsfonts}
\usepackage{txfonts}
\usepackage{amssymb}
\usepackage{graphicx}
\usepackage{subeqnarray}
\usepackage{xcolor} 
 
\newcommand{\ket}[1]{|#1\rangle}
\newcommand{\bra}[1]{\langle #1|}
\newcommand{\Tr}{\text{Tr}}
\begin{document}
\today
\title{Robust paths to realize nonadiabatic holonomic gates}
\author{G. F. Xu}
\affiliation{Department of Physics, Shandong University, Jinan
250100, China}
\affiliation{Department of Physics and Astronomy, Uppsala University,
Box 516, Se-751 20 Uppsala, Sweden}
\author{P. Z. Zhao}
\affiliation{Department of Physics, Shandong University, Jinan
250100, China}
\author{D. M. Tong}
\email{ e-mail: tdm@sdu.edu.cn}
\affiliation{Department of Physics, Shandong University, Jinan
250100, China}
\author{Erik Sj\"oqvist}
\email{e-mail: erik.sjoqvist@physics.uu.se}
\affiliation{Department of Physics and Astronomy, Uppsala University,
Box 516, Se-751 20 Uppsala, Sweden}

\begin{abstract}
To realize one desired nonadiabatic holonomic gate, various equivalent evolution paths can
be chosen. However, in the presence of errors, these paths become inequivalent. In this paper,
we investigate the difference of these evolution paths in the presence of systematic Rabi 
frequency errors and aim to find paths with optimal robustness to realize one-qubit nonadiabatic 
holonomic gates. We focus on three types of evolution paths in the $\Lambda$ system: paths 
belonging to the original two-loop scheme [New J. Phys. {\bf 14}, 103035 (2012)], the single-loop
multiple-pulse scheme [Phys. Rev. A {\bf 94}, 052310 (2016)], and the off-resonant single-shot
scheme [Phys. Rev. A {\bf 92}, 052302 (2015); Phys. Lett. A {\bf 380}, 65 (2016)]. Whereas both
the single-loop multiple-pulse and single-shot schemes aim to improve the robustness of the
original two-loop scheme by shortening the exposure to decoherence, we here find that the
two-loop scheme is more robust to systematic errors in the Rabi frequencies. More
importantly, we derive conditions under which the resilience to this kind of error can be
optimized, thereby strengthening the robustness of nonadiabatic holonomic gates.
\end{abstract}
\maketitle
\date{\today}

\section{Introduction}

Quantum computers process information encoded in quantum systems and use logical
operations different from those in Boolean logic. In this way, quantum computation
may be able to solve problems faster than in classical computation. Generally,
to achieve the advantages of quantum computation, a universal set of quantum gates with
sufficiently high fidelities needs to be realized. However, such a realization is difficult because
of the detrimental effects of errors, such as decoherence and parameter noise. To
overcome this problem, different forms of error resilient strategies have been developed,
such as error correcting codes \cite{knill97}, decoherence-free subspaces and subsystems
\cite{lidar98,zanardi00}, and techniques based on geometrical and topological phases \cite{zanardi99,pachos12}.

Holonomic quantum computation exploits Abelian or non-Abelian geometric phases to
implement a universal set of quantum gates \cite{zanardi99,zhu02,Sjoqvist}. These gates
only depend on evolution paths, but not on evolution details, and thus may have robust
advantages over dynamical gates in the presence of noise \cite{Chiara,Solinas,Zhu2005,Florio,Parodi,Lupo,Thomas,Solinas1,Johansson,Jing2017}.
Due to its high-speed nature,
nonadiabatic holonomic quantum computation proposed in Ref.~\cite{Sjoqvist} has received
considerable attention. So far, many schemes of realizing nonadiabatic holonomic gates have
been put forward for various physical systems
\cite{Xu,Abdumalikov,Feng,Mousolou,Zhang1,Arroyo,Zu,Xu2,Liang,
Zhou,Xue,Pyshkin,Song,Xue1,Wang,Herterich,Sun,Xu3,Sjoqvist2,Zhao,Sekiguchi2017,Zhou2017,Li2017}. Particularly, nonadiabatic holonomic quantum computation has been experimentally demonstrated 
with circuit QED \cite{Abdumalikov}, NMR \cite{Feng,Li2017}, and nitrogen-vacancy centers in 
diamond \cite{Arroyo,Zu,Sekiguchi2017,Zhou2017}.

To realize one desired nonadiabatic holonomic gate in the standard $\Lambda$
setting, various types of evolution paths can be chosen. These evolution paths are
equivalent in the ideal case, but behave differently in the presence of errors. To improve
further on the robustness of nonadiabatic holonomic gates, investigating the robustness of
these paths is therefore of importance. In this paper, we examine the evolution paths
in the presence of systematic errors in the Rabi frequencies, which is known as the main
error source when the evolution period is shorter than the coherence time \cite{low14,ivanov15}.
We focus on the types of evolution paths proposed in Refs.~\cite{Sjoqvist,Xu3,Sjoqvist2,Herterich}. 
These investigated evolution paths are based on three-level systems driven by laser fields and 
the Rabi frequency errors occur when the amplitudes of the driving fields deviate from their desired 
values by unknown fractions. We aim to find paths with optimal robustness to realize one-qubit 
nonadiabatic holonomic gates. We derive conditions under which the resilience to Rabi frequency 
errors can be optimized with respect to the evolution path belonging to these implementations. 
In this way, we strengthen the robustness of nonadiabatic holonomic quantum computation.

\section{Paths affected by systematic errors}

\subsection{Two-loop scheme}
We start by investigating evolution paths belonging to the original two-loop scheme
\cite{Sjoqvist}. Consider a three-level system whose states are $\ket{0}$, $\ket{1}$, and
$\ket{e}$, where $\ket{0}$ and $\ket{1}$ are used as logical one-qubit states and $\ket{e}$ is used as an
ancillary state. The transitions $\ket{0}\leftrightarrow\ket{e}$ and $\ket{1}\leftrightarrow\ket{e}$
are driven by two resonant laser pulses with identical envelope. By transforming
to the rotating frame, the Hamiltonian of the system reads
\begin{eqnarray}
H(t) & = & \Omega(t)\left[ e^{i\varphi}\big(\cos\frac{\theta}{2}\ket{0}
+\sin\frac{\theta}{2}{e^{i\psi}}\ket{1}\big)\bra{e}+\textrm{H.c.}\right]
\nonumber \\
 & = & \Omega(t) \left[ e^{i\varphi} \ket{b} \bra{e} + e^{-i\varphi} \ket{e} \bra{b} \right] ,
\label{H}
\end{eqnarray}
where $\ket{b}=\cos(\theta/{2})\ket{0} + \sin(\theta/{2}) e^{i\psi}\ket{1}$
is the bright state and we have neglected rapidly oscillating counter-rotating terms (rotating wave
approximation). Here, $\Omega(t)$ is the pulse envelope, $\theta=2\arctan[\Omega_1(t)/\Omega_0(t)]$,
and $\varphi$ and $\psi$ respectively represent the total and relative phases, with $\Omega_0(t)$
and $\Omega_1(t)$ being Rabi frequencies of the two laser fields. The laser parameters
$\theta,\psi$, and $\varphi$ are kept constant during each pulse pair. To realize the desired
holonomic one-qubit gate, two evolution loops $1$ and $2$ are needed. The duration $T_{\nu}$,
of each loop $\nu\in\{1,2\}$  satisfies $\int_0^{T_{\nu}}\Omega(t)dt=\pi$. Correspondingly, the
time evolution operator reads
\begin{eqnarray}
U_{\nu}=-\ket{e}\bra{e}-\vec{n}_{\nu}\cdot\vec{\sigma},
\end{eqnarray}
where $\vec{n}_{\nu}=(\sin\theta_{\nu}\cos\psi_{\nu}, \sin\theta_{\nu}\sin\psi_{\nu}, \cos\theta_{\nu})$
is a unit vector in ${R}^3$ and $\vec{\sigma}=(\sigma_x, \sigma_y, \sigma_z)$ is the standard
vector Pauli operator acting on states $\ket{0}$ and $\ket{1}$. The parameters $\theta_{\nu}$,
$\psi_{\nu}$, and $\varphi_{\nu}$ determine the loop $\nu$.  Note that the gate $U_{\nu}$ is
independent of the total phase $\varphi_{\nu}$ in this ideal case. The two-loop holonomic gate
is realized by sequentially implementing $U_1$ and $U_2$, yielding
\begin{eqnarray}
U = U_2 U_1 & = & \ket{e} \bra{e} + \vec{n}_1 \cdot \vec{n}_2 (\ket{0}\bra{0}+\ket{1}\bra{1})
\nonumber \\
 & & -i (\vec{n}_1\times\vec{n}_2)\cdot\vec{\sigma}
\nonumber \\
 & \equiv & \ket{e} \bra{e} + R(\vartheta,\vec{m}) ,
\label{ul}
\end{eqnarray}
where
\begin{eqnarray}
R(\vartheta,\vec{m})=\exp(i\vartheta\vec{m}\cdot\vec{\sigma})
\end{eqnarray}
In the above, $2\vartheta \in [0, \pi]$ is the rotation angle and ${\vec{m}}=\vec{n}_2\times\vec{n}_1/|\vec{n}_2\times\vec{n}_1|$ (provided $|{\vec{n}_2}\times{\vec{n}_1}|\neq0$) is the rotation axis of
the one-qubit holonomic gate acting on the computational subspace spanned by
$\{ \ket{0},\ket{1} \}$. It can be shown that the two-loop gate $U$ is of holonomic
nature, i.e., that all dynamical contributions to the gate vanish \cite{Sjoqvist}.

It is noteworthy that all unit vector pairs $\vec{n}_1,\vec{n}_2$ with the same relative angle
in a given plane can be used to realize one desired gate. This means that an infinite number
of evolution paths are equivalent in the ideal case. These evolution paths may behave
differently in the presence of errors. Here, we investigate their difference and aim to optimize
their robustness to systematic errors.

For Hamiltonian $H(t)$ as realized by laser fields, the main source of systematic errors is in the 
Rabi frequencies \cite{low14,ivanov15}. This kind of errors occurs when the amplitude of the 
driving field deviates from its desired value by an unknown fraction. Since this kind of errors 
is closely related to the pulse amplitude, it is also called systematic amplitude error \cite{low14}. 
If one analyzes the evolution operator generated by a Rabi-frequency-error-affected Hamiltonian, 
the rotation angle will be found to deviate from its desired value. Then the Rabi frequency error 
can also be seen as a rotation error \cite{ivanov15}.

Since the two laser fields are typically calibrated by the same standard, it
is experimentally natural to assume that they experience the same Rabi frequency error.
In this case, the Hamiltonian $H(t)$ turns into
\begin{eqnarray}
H^{\prime}(t) = (1+\epsilon) \Omega(t) \left[ e^{i\varphi} \ket{b} \bra{e} +
e^{-i\varphi} \ket{e} \bra{b} \right] ,
\end{eqnarray}
where $\epsilon$ is the unknown fraction that satisfies $\mid\epsilon\mid\ll1$.
By using the expression of $H^{\prime}(t)$, we obtain the error-affected two-loop gate
\begin{eqnarray}
U^{\prime} & = & U_2\exp[-i\epsilon\pi(e^{i\varphi_{2}}\ket{b_2}\bra{e}+\textrm{H.c.})]
\nonumber\\
 & & \times \exp[-i\epsilon\pi(e^{i\varphi_{1}}\ket{b_1}\bra{e}+\textrm{H.c.})]U_1
\end{eqnarray}
with the bright state $\ket{b_{\nu}}=\cos({\theta_{\nu}}/{2})\ket{0} +
\sin({\theta_{\nu}}/{2}){e^{i\psi_{\nu}}}\ket{1}$.

To quantify the robustness of the holonomic gates, we use the fidelity \cite{Wang2009}
\begin{eqnarray}
F(V) = \frac{\left| \Tr (V^\dag{V_e})\right|}{\Tr (V^\dag{V})},
\end{eqnarray}
where $V$ and $V_e$ are the desired and error-affected evolution operators, respectively.
By substituting $U$ and $U^{\prime}$ into $F$, we obtain
\begin{eqnarray}
F(U) & = & \frac{1}{3} \left| \Tr \big\{\exp \left[ -i\epsilon\pi(e^{i\varphi_{2}}\ket{b_2}
\bra{e}+\textrm{H.c.})
\right] \right.
\nonumber \\
 & & \left. \times \exp \left[-i\epsilon\pi(e^{i\varphi_{1}}\ket{b_1}\bra{e}+\textrm{H.c.})
 \right]\big\} \right| .
\end{eqnarray}
To simplify, we use the Baker-Campbell-Hausdorff (BCH) formula
$\ln(e^Ae^B)={A+B+\frac{1}{2}[A,B]+\cdots}$. Since $\mid\epsilon\mid\ll1$, we only
keep terms to first- and second-order in $\epsilon$, yielding the fidelity
\begin{eqnarray}
F&=&\frac{1}{3} \left| \Tr \big\{\exp[-i\epsilon\pi(e^{i\varphi_{2}}\ket{b_2}\bra{e}
+e^{i\varphi_{1}}\ket{b_1}\bra{e}+\textrm{H.c.})  \right.
\nonumber \\
& & \left. +(AB-BA)/2 ]\big\}\right| ,
\end{eqnarray}
where $A=-i\epsilon\pi(e^{i\varphi_{2}}\ket{b_2}
\bra{e}+\textrm{H.c.})$, $B=-i\epsilon\pi(e^{i\varphi_{1}}\ket{b_1}\bra{e}+\textrm{H.c.})$ and thus $(AB-BA)/2$ represents the second-order terms in $\epsilon$.
It can be verified that
\begin{eqnarray}
e^{i\varphi_{2}}\ket{b_2}=\cos\frac{\eta}{2}{e^{i(\phi_b + \varphi_1)}}\ket{b_1}
+\sin\frac{\eta}{2}{e^{i\phi_d}}\ket{d_1},
\label{relation}
\end{eqnarray}
where the dark state $\ket{d_1}=\sin({\theta_1}/{2})\ket{0} -\cos({\theta_1}/{2}){e^{i\psi_{1}}}
\ket{1}$, $\eta$ is the angle between the state vectors $\ket{b_1}$ and $\ket{b_2}$ in the
Bloch sphere representing the one-qubit subspace, and $\phi_b$ and $\phi_d$ are phases.
In Eq. (\ref{relation}), we use the states $e^{i\varphi_1}\ket{b_1}$ and $\ket{d_1}$ as the
orthonormal basis to express the state   $e^{i\varphi_2}\ket{b_2}$. In this case, $\phi_b$
and $\phi_d$ are the decomposition phases with respect to the corresponding basis states.
By using Eq. (\ref{relation}), we rewrite the fidelity as
\begin{eqnarray}
F&=&\frac{1}{3} \left| \Tr\big\{\exp[-i\epsilon\pi{x}(\ket{b_{12}}\bra{e}
+\ket{e}\bra{b_{12}}) \right.
\nonumber \\
& & \left. +(AB-BA)/2]\big\}\right|
\end{eqnarray}
with
\begin{eqnarray}
\ket{b_{12}} & = & \left[ \left(1+\cos(\eta/2)e^{i\phi_b} \right) e^{i\varphi_{1}}\ket{b_1} \right.
\nonumber \\
 & & \left. + \sin(\eta/2) e^{i\phi_d} \ket{d_1} \right] /x,
\nonumber \\
x & = & [2+2\cos(\eta/2)\cos{\phi_b}]^{1/2} .
\end{eqnarray}
By expanding the exponential operator, we can rewrite the fidelity as
\begin{eqnarray}
F&=&\frac{1}{3} \left| \Tr\big\{\sum_{n=0}^{+\infty}[-i\epsilon\pi{x}(\ket{b_{12}}\bra{e}
+\ket{e}\bra{b_{12}}) \right.
\nonumber \\
& & \left. +(AB-BA)/2]^n\big\}\right|
\end{eqnarray}
It is known that $(AB-BA)/2$ represents terms of the second-order and its trace $\Tr[(AB-BA)/2]=0$. By using that $\langle e \ket{b_{12}} = 0$,
we also find $\Tr (\ket{e}\bra{b_{12}}+\ket{b_{12}}\bra{e})=0$ and $(\ket{e}\bra{b_{12}}+\ket{b_{12}}\bra{e})^2=\ket{e}\bra{e}+\ket{b_{12}}\bra{b_{12}}$.
Thus the fidelity can be further simplified to
\begin{eqnarray}
F=1-\frac{2}{3} \left( 1+\cos\frac{\eta}{2}\cos{\phi_b} \right)\pi^2\epsilon^2
\end{eqnarray}
up to the second-order in $\epsilon$.

In the above, we have derived the fidelity for various paths in the two-loop scheme. To find
which path is more robust, it is convenient to rewrite the fidelity in terms of holonomic gate
$R(\vartheta,\vec{m})$. If we focus on the above logical gate, the parameters $\vartheta$
and ${\vec{m}}$ are fixed and the corresponding fidelity reads
\begin{eqnarray}
F_R = 1-\frac{2}{3} \left( 1+\cos\frac{\vartheta}{2}\cos{\phi_b} \right)\pi^2\epsilon^2.
\end{eqnarray}
It is noteworthy that the parameter $\phi_b$ can be controlled by the relative phase shift
between the two pulse pairs, which shows the difference of the
evolution paths to realize the holonomic gate $R(\vartheta,\vec{m})$ in the presence of errors.
Since $\vartheta\in [0, \pi/2]$, we should set
\begin{eqnarray}
\phi_b=\pi
\end{eqnarray}
in order to minimize the error $\epsilon$ dependence, yielding
\begin{eqnarray}
F_R=1-\frac{2}{3} \left( 1-\cos\frac{\vartheta}{2} \right) \pi^2\epsilon^2.
\end{eqnarray}
From the above equation, we can see that the fidelity only depends on the rotation
angle $\vartheta$, but not on the rotation axis $\vec{m}$. Furthermore, $F_R$ decreases
when the rotation angle $\vartheta$ increases.

\subsection{Single-loop multiple-pulse scheme}

For the single-loop multiple-pulse scheme \cite{Herterich}, the three-level system
is again driven by pairs of resonant laser pulses with identical envelop. Thus, the
Hamiltonian $\tilde{H}(t)$ in the ideal single-loop multiple-pulse scheme has the
same form as $H(t)$.

In its simplest form, the single-loop multiple-pulse gate is realized by splitting the whole
evolution time $T$ into two segments and the control Hamiltonians for these two segments
have different total phases denoted as $\varphi$ and $\varphi^{\prime}$, respectively. For
the first segment, the total phase is equal to $\varphi$ and its evolution period satisfies
$\int_0^{\tau} \Omega (t)dt = \pi/2$, where $\tau$ is the intermediate time. The
corresponding evolution operator reads
\begin{eqnarray}
U_{\varphi}=-i \left( e^{i\varphi}\ket{{b}}\bra{e}+e^{-i\varphi}\ket{e}\bra{{b}} \right)
+\ket{{d}}\bra{{d}},
\end{eqnarray}
with the dark state $\ket{{d}}=\sin({\theta}/{2})\ket{0}-\cos({\theta}/{2}){e^{i\psi}}\ket{1}$.
For the second segment, the total phase is changed from $\varphi$ to $\varphi^{\prime}$
and the time period satisfies $\int_{\tau}^{T} \Omega(t)dt = \pi/2$. The resulting evolution
operator reads
\begin{eqnarray}
U_{\varphi^{\prime}}=-i\left( e^{i\varphi^{\prime}}\ket{{b}}\bra{e}
+e^{-i\varphi^{\prime}}\ket{e}\bra{{b}} \right) + \ket{d}\bra{d}.
\end{eqnarray}
By sequentially implementing $U_{\varphi}$ and $U_{\varphi^{\prime}}$, the single-loop
multiple-pulse gate
\begin{eqnarray}
\tilde{U}=-e^{i(\varphi-\varphi^{\prime})}\ket{e}\bra{e}
-e^{-i(\varphi-\varphi^{\prime})}\ket{{b}}\bra{{b}}
+\ket{{d}}\bra{{d}}
\end{eqnarray}
is realized. It can be verified that the single-loop multiple-pulse gate is of nonadiabatic
holonomic nature.

In the presence of Rabi frequency errors, the Hamiltonian can be written as
\begin{eqnarray}
\tilde{H}^{\prime}(t) & = & (1+\epsilon) \Omega (t) \left( e^{i\varphi}\ket{{b}}\bra{e}
+e^{-i\varphi}\ket{e}\bra{{b}} \right) ,
\end{eqnarray}
where $\epsilon$ is the unknown fraction. The single-loop multiple-pulse gate
$\tilde{U}$ turns into
\begin{eqnarray}
\tilde{U}^{\prime} & = & U_{\varphi^{\prime}} \exp \left[ -i\epsilon\frac{\pi}{2}
(e^{i\varphi^{\prime}}\ket{{b}}\bra{e}+e^{-i\varphi^{\prime}}\ket{e}\bra{{b}}) \right]
\nonumber\\
 & & \times \exp \left[ -i\epsilon\frac{\pi}{2}(e^{i\varphi}\ket{{b}}\bra{e}
+e^{-i\varphi}\ket{e}\bra{{b}}) \right] U_{\varphi}.
\end{eqnarray}
By using the expressions of $\tilde{U}$ and $\tilde{U}^{\prime}$, we obtain the gate fidelity
\begin{eqnarray}
\tilde{F} & = & \frac{1}{3} \left| \Tr \left\{ \exp \left[ -i\epsilon\frac{\pi}{2}
(e^{i\varphi^{\prime}}\ket{{b}}\bra{e}+e^{-i\varphi^{\prime}}\ket{e}\bra{{b}}) \right] \right. \right.
\nonumber\\
 & & \left. \left. \times \exp \left[ -i\epsilon\frac{\pi}{2}(e^{i\varphi}\ket{{b}}\bra{e}
+e^{-i\varphi}\ket{e}\bra{{b}}) \right] \right\} \right|.
\end{eqnarray}
The BCH relation yields
\begin{eqnarray}
\tilde{F} &=& \frac{1}{3} \left| \Tr \left\{ \exp \left[ -i\epsilon\frac{\pi}{2}(e^{i\varphi}
\ket{{b}}\bra{e} + e^{i\varphi^{\prime}}\ket{{b}}\bra{e}+\textrm{H.c.}) \right. \right. \right.
\nonumber \\
& & \left. \left. \left. +(\tilde{A}\tilde{B}-\tilde{B}\tilde{A})/2 \right] \right\} \right| ,
\end{eqnarray}
where operators $\tilde{A}=-i\epsilon\pi
(e^{i\varphi^{\prime}}\ket{{b}}\bra{e}+e^{-i\varphi^{\prime}}\ket{e}\bra{{b}})/2$ and $\tilde{B}=-i\epsilon\pi(e^{i\varphi}\ket{{b}}\bra{e}
+e^{-i\varphi}\ket{e}\bra{{b}})/2$, and we have kept terms up to second-order in $\epsilon$, since $| \epsilon | \ll 1$.
By evaluating the trace, the gate fidelity $\tilde{F}$ can be written as
\begin{eqnarray}
\tilde{F}=1-\frac{1}{6} \left[ 1+\cos \left( \varphi-\varphi^{\prime} \right) \right] \pi^2\epsilon^2.
\end{eqnarray}

Similar to the case of the two-loop scheme, we rewrite the fidelity $\tilde{F}$ in terms
of the logical gate $R(\vartheta, \vec{m})$, yielding
\begin{eqnarray}
\tilde{F}_R=1-\frac{1}{6} \left[ 1-\cos \left( 2\vartheta \right) \right]\pi^2\epsilon^2.
\end{eqnarray}
The above equation shows that the fidelity only depends on the rotation angle $\vartheta$,
but not on the rotation axis $\vec{m}$, and it decreases when $\vartheta$ increases.

\subsection{Single-shot scheme}
We finally investigate paths belonging to the single-shot scheme originally proposed in
Refs.~\cite{Xu3,Sjoqvist2}. Unlike the two-loop and single-loop multiple-pulse schemes, the
transitions $\ket{0}\leftrightarrow\ket{e}$ and $\ket{1}\leftrightarrow\ket{e}$ of the
three-level system are driven by off-resonant laser pulses in the single-shot scheme.
Explicitly, the two laser pulses should have the same detuning and be square shaped.
In the rotating frame, the Hamiltonian of the system reads
\begin{eqnarray}
\hat{H}(t) = -\Delta\ket{e}\bra{e}+\sum_{j=0}^1\left( \Omega_je^{i\beta_j} \ket{j}\bra{e} +
\textrm{H.c.} \right),
\end{eqnarray}
where $\Delta$ is the detuning, ${\Omega}_j$ is the Rabi frequency, and
$\beta_j$ is the phase. We introduce new parameters $\Omega,\alpha$, and $\gamma$
according to
\begin{eqnarray}
\left\{\begin{array}{c}
\Delta=-2{\Omega}\sin\gamma, \\
{\Omega}_0={\Omega}\cos\alpha\cos\gamma,\\
{\Omega}_1={\Omega}\sin\alpha\cos\gamma,
\end{array}\right. \label{conditionsingle}
\end{eqnarray}
where ${\Omega}$ can be seen as the norm of the vector $(\Delta/2, \Omega_0, \Omega_1)$.
As a result, the Hamiltonian $\hat{H}(t)$ can be rewritten as
\begin{eqnarray}
\hat{H}(t) & = & \Omega \sin\gamma (\ket{e}\bra{e}+\ket{b}\bra{b}) +
\Omega [\cos\gamma(\ket{b}\bra{e}
\nonumber \\
 & & +\ket{e}\bra{b}) +
\sin\gamma(\ket{e}\bra{e}-\ket{b}\bra{b})]
\end{eqnarray}
with the bright state $\ket{b}=\cos\alpha{e^{i\beta_0}}\ket{0}+\sin\alpha{e^{i\beta_1}}\ket{1}$.
By choosing the evolution time such that ${\Omega}{T}=\pi$, the single-shot holonomic gate is
realized and reads
\begin{eqnarray}
\hat{U}=e^{i\zeta}(\ket{e}\bra{e}+\ket{b}\bra{b})+\ket{d}\bra{d},
\end{eqnarray}
where phase $\zeta=\pi-\pi\sin\gamma$ and the dark state $\ket{d}$ is orthogonal to both
$\ket{e}$ and $\ket{b}$. In the above, the rotation angle of the qubit gate is determined by
$\sin\gamma$, while the rotation axis is determined by $\ket{b}$ and $\ket{d}$. It can
be verified that the single-shot gate has a nonadiabatic holonomic feature \cite{Xu3,Sjoqvist2}.

The single-shot scheme is realized by varying the detuning of the two laser fields. When the
detuning is small, i.e., $|\cos \gamma|$ is close to unity, the Rabi frequency error model
can appropriately describe the systematic errors of the system. This is because the Hamiltonian
under the condition of small detuning is nearly the same as the Hamiltonian of the two-loop
and single-loop multi-pulse schemes. In the regime where $|\sin \gamma|$ is close to
unity, the detuning is large relative the Rabi frequencies $\Omega_j$, as can be seen in
Eq.~(\ref{conditionsingle}).
Thus, in this case the influence of the Rabi frequency errors $\epsilon\Omega_j$ on the gate
is weak. Meanwhile, several control problems start appearing. For example, the rotation wave
approximation may not work any longer because of small run time to realize cyclic
evolution in this regime \cite{Sjoqvist2}. In this case, the Rabi frequency errors are no
longer the dominant error source. On the other hand, if we artificially consider the Rabi
frequency errors only, the fidelity would be higher than that obtained by including the dominant
errors. In the next section, we will see that even though these higher resulting fidelities
are used, the paths belonging to the single-shot scheme are still the least robust. Since
our aim is to find the paths with optimal robustness, we can simply use the Rabi frequency
error model to compare the single-shot scheme with the two-loop and multiple-pulse schemes.

In the presence of Rabi frequency errors, the Hamiltonian $\hat{H}(t)$ turns into
\begin{eqnarray}
\hat{H}^{\prime}(t) & = & 2{\Omega}\sin\gamma\ket{e}\bra{e}
\nonumber \\
 & & + (1+\epsilon) \Omega \cos \gamma (\ket{b}\bra{e} +\ket{e}\bra{b}) ,
\end{eqnarray}
where $\epsilon$ is the error fraction. The Hamiltonian $\hat{H}^{\prime}(t)$
turns the desired single-shot gate into
\begin{eqnarray}
\hat{U}^{\prime}=e^{-i\pi\sin\gamma(\ket{e}\bra{e}+\ket{b}\bra{b})}
e^{-i\lambda\pi\sigma_\epsilon}+\ket{d}\bra{d}
\end{eqnarray}
with the parameter $\lambda=[(1+\epsilon)^2\cos^2\gamma + \sin^2\gamma]^{1/2}$
and Pauli operator $\sigma_\epsilon = [(1+\epsilon)\cos\gamma(\ket{b}\bra{e} +
\ket{e}\bra{b} +\sin\gamma(\ket{e}\bra{e}-\ket{b}\bra{b})]/\lambda$. The fidelity of the
unitary operators $\hat{U}$ and $\hat{U}^{\prime}$ reads
\begin{eqnarray}
\hat{F} = \frac{1}{3} \left| \Tr \left( \ket{d}\bra{d}-e^{-i\lambda\pi\sigma_\epsilon}
\right) \right|.
\end{eqnarray}
Considering that $\sigma_\epsilon^2=\ket{e}\bra{e}+\ket{b}\bra{b}$ and
$\Tr(\sigma_\epsilon)=0$, the fidelity $\hat{F}$ can be further simplified into
\begin{eqnarray}
\hat{F} = \frac{1}{3} \left| \Tr \left[ \ket{d}\bra{d} - \cos (\lambda \pi) (\ket{e}\bra{e} +
\ket{b}\bra{b}) \right] \right| .
\end{eqnarray}
Since $| \epsilon | \ll 1$, we can ignore higher-order terms in $\epsilon$ and the
gate fidelity $\hat{F}$ reads
\begin{eqnarray}
\hat{F} = 1 - \frac{1}{3} \pi^2 \epsilon^2 \cos^4\gamma.
\end{eqnarray}
By expressing $\gamma$ in terms of the gate rotation angle $\vartheta$, we obtain the fidelity
\begin{eqnarray}
\hat{F}_R=1-\frac{16}{3}\vartheta^2 \left( 1-\vartheta/\pi \right)^2\epsilon^2.
\end{eqnarray}
of the logical gate $R(\vartheta, \vec{m})$. We see that the fidelity only depends on the
rotation angle $\vartheta$, but not on the rotation axis $\vec{m}$. It can be verified that
the fidelity decreases when the rotation angle increases.

\subsection{Robustness of the paths}
We now examine the robustness of the selected evolution paths by comparing the fidelities
$F_R$, $\tilde{F}_R$, and $\hat{F}_R$ for a given gate realization $R(\vartheta,\vec{m})$.
As was shown previously, these fidelities depend on the rotation angle $\vartheta$, but
are independent of the rotation axis $\vec{m}$. Explicitly, we may write
\begin{eqnarray}
F_R & = & 1-\frac{1}{3}\pi^2\epsilon^2 f_1,
\nonumber \\
\tilde{F}_R & = & 1-\frac{1}{3}\pi^2\epsilon^2 f_2,
\nonumber \\
\hat{F}_R & = & 1-\frac{1}{3}\pi^2\epsilon^2 f_3,
\end{eqnarray}
where $f_1=2-2\cos(\vartheta/2)$, $f_2=(1-\cos2\vartheta)/2$, and
$f_3=16\vartheta^2(1-\vartheta/\pi)^2/\pi^2$ describe the error of the
two-loop \cite{Sjoqvist}, single-loop multiple-pulse \cite{Herterich}, and
single-shot \cite{Xu3,Sjoqvist2} schemes, respectively. Thus, we can address our problem by
comparing the functions $f_1$, $f_2$, and $f_3$ instead of the fidelities $F_R$, $\tilde{F}_R$,
and $\hat{F}_R$. The benefit with this comparison is that the dependence on the value of
the error, i.e., $\epsilon$, is excluded.

\begin{figure}[htbp]
\begin{center}
\includegraphics[width=8.5cm, height=4.5cm]{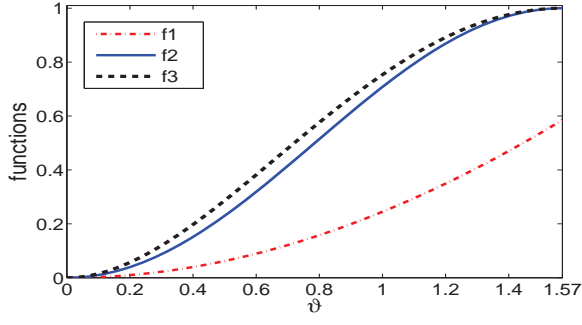}
\end{center}
\caption{(color online). Functions $f_1$, $f_2$, and $f_3$ as functions of the rotation angle
$\vartheta$ of the one-qubit holonomic gate. Here, $f_1=2-2\cos(\vartheta/2)$, $f_2=(1-\cos2\vartheta)/2$, $f_3=16\vartheta^2(1-\vartheta/\pi)^2/\pi^2$, and
$\vartheta\in [0, \pi/2]$. The functions $f_1$, $f_2$, and $f_3$ describe the error
of the two-loop \cite{Sjoqvist}, single-loop multiple-pulse \cite{Herterich}, and
single-shot \cite{Xu3,Sjoqvist2} schemes, respectively.}
\label{fig1}
\end{figure}

Figure \ref{fig1} shows $f_1$, $f_2$, and $f_3$ as functions of the rotation angle $\vartheta$.
From this figure, we see that $f_1$, $f_2$, and $f_3$ all increase with increasing rotation angle
$\vartheta$. We also see that the function $f_1$ is always smaller than the other two functions
and the differences $f_1-f_2$ and $f_1-f_3$ increase with increasing $\vartheta$.

The observations of Fig.~\ref{fig1} show that the most robust path to realize the gate
$R(\vartheta, \vec{m})$ belongs to the two-loop scheme. We have derived the condition to optimize the performance of the gate $R(\vartheta, \vec{m})$. Specifically, the parameters should satisfy the condition
\begin{eqnarray}
\phi_b = \pi. \label{conditionpi}
\end{eqnarray}
The above fact helps us to reconsider the advantages of the two-loop, the single-loop
multiple-pulse, and the single-shot schemes.

It has been shown that the parameter $\phi_b$ is particularly important in optimizing the robustness
of the two-loop gates. In fact, if the parameter $\phi_b$ does not satisfy the condition in
Eq. (\ref{conditionpi}), for example, by being made to equal zero, one will find that the fidelity
of the path of the two-loop scheme becomes less robust than the other investigated schemes,
which shows the necessity of the choice of the robust path.

\section{Conditions to realize a more robust path with relative difference}
In the preceding section, we have examined the robustness of paths to implement a given
nonadiabatic holonomic one-qubit gate in the presence of Rabi frequency errors in the
$\Lambda$ setting. We have assumed that the two laser pulses experience the same Rabi
frequency error, motivated by the experimentally natural assumption that the pulses are
calibrated by the same standard. We now relax this assumption by introducing a relative
difference between the two Rabi frequencies. From the preceding section, we already know
which kinds of paths are more robust when the relative difference is zero. Here we assume
the relative difference is small, and in this case the robust advantages of the chosen paths
with optimal robustness prevail. As already known, the chosen paths belong to the two-loop
scheme. So, it is sufficient to consider only the two-loop scheme in the following analysis.

The Hamiltonian of the system can now be written as
\begin{eqnarray}
H^{\prime\prime}(t) & = & \Omega(t) \left\{ e^{i\varphi} \left[ (1+\epsilon_0)
\cos\frac{\theta}{2}\ket{0} \right. \right.
\nonumber\\
 & & \left. \left. + (1+\epsilon_1) \sin\frac{\theta}{2}{e^{i\psi}}\ket{1} \right] \bra{e} +
\textrm{H.c.} \right\},
\label{hpp}
\end{eqnarray}
where $\epsilon_0$ and $\epsilon_1$ are unknown fractions of the two Rabi frequencies
associated with the pulse pair. According to this Hamiltonian, the desired nonadiabatic
holonomic one-qubit gate is modified into
\begin{eqnarray}
U^{\prime\prime} & = & U_2^{\prime} \exp \left[ -i\delta_2\pi
\left( e^{i\varphi_2}\ket{{b}_2^{\prime}}\bra{e} + \textrm{H.c.} \right) \right]
\nonumber\\
 & & \exp \left[-i\delta_1\pi \left( e^{i\varphi_1}\ket{{b}_1^{\prime}}\bra{e}+\textrm{H.c.}
\right) \right]
U_1^{\prime} ,
\end{eqnarray}
where
\begin{eqnarray}
U_{\nu}^{\prime} & = & -\ket{e}\bra{e} - \ket{{b}_{\nu}^{\prime}} \bra{{b}_{\nu}^{\prime}} +
\ket{{d}_{\nu}^{\prime}} \bra{{d}_{\nu}^{\prime}} ,
\nonumber \\
\delta_{\nu} & = & [ (1+\epsilon_0)^2\cos^2(\theta_{\nu}/2)
 \nonumber \\
 & & + (1+\epsilon_1)^2 \sin^2(\theta_{\nu}/2)]^{1/2}-1,
\end{eqnarray}
where $\nu = 1,2$. Here, $\ket{{b}_{\nu}^{\prime}}=\cos(\theta_{\nu}^{\prime}/2)\ket{0} +
\sin(\theta_{\nu}^{\prime}/2) e^{i\psi_{\nu}}\ket{1}$ and $\ket{{d}_{\nu}^{\prime}} =
\sin(\theta_{\nu}^{\prime}/2)\ket{0} -\cos(\theta_{\nu}^{\prime}/2){e^{i\psi_{\nu}}}\ket{1}$
are the bright and dark states, respecteively, with the error dependent angle
$\theta_{\nu}^{\prime} = 2\arctan[\tan(\theta_{\nu}/2) (1+\epsilon_1)/(1+\epsilon_0)]$.
By using the expressions for $U$ and $U^{\prime\prime}$, we obtain the fidelity
\begin{eqnarray}
F^{\prime\prime} & = & \frac{1}{3} \left| \Tr \left\{ U_1^{\prime} U_1^{\dag} U_2^{\dag}
U_2^{\prime} \exp\left[ -i\delta_2\pi \left( e^{i\varphi_{2}}\ket{b_2^{\prime}}\bra{e} +
\textrm{H.c.} \right) \right] \right. \right.
\nonumber\\
 & & \left. \left. \exp \left[ -i\delta_1 \pi \left( e^{i\varphi_{1}}\ket{b_1^{\prime}}\bra{e} +
\textrm{H.c.} \right) \right] \right\} \right| .
\end{eqnarray}
To simplify this, we first consider the operator $U_1^{\prime} U_{1}^{\dag}U_{2}^{\dag}
U_2^{\prime}$ which, in the following, is denoted as $U_{12}$ for convenience. We have
\begin{eqnarray}
U_{12}=\ket{e}\bra{e}
+\exp\big(-i\theta_{22}\sigma_{\psi_{2}+\frac{\pi}{2}}\big)
\exp\big(i\theta_{11}\sigma_{\psi_{1}+\frac{\pi}{2}}\big),
\end{eqnarray}
where $\theta_{\nu\nu}=\theta_{\nu}-\theta_{\nu}^{\prime}$ and $\sigma_{\psi_{\nu}+\pi/2}=\cos(\psi_{\nu}+\pi/2)\sigma_x
+\sin(\psi_{\nu}+\pi/2)\sigma_y$, with $\sigma_x$ and $\sigma_y$ being the Pauli
operators acting on the subspace spanned by $\{\ket{0}, \ket{1}\}$. It can be verified
that $| \theta_{\nu\nu} | \ll 1$. Thus, we can use the BCH formula to further simplify
the operator $U_{12}$. By preserving the first-order terms, we obtain
\begin{eqnarray}
U_{12}=\ket{e}\bra{e}
+\exp\big[i(\theta_{11}\sigma_{\psi_{1}+\frac{\pi}{2}}-\theta_{22}
\sigma_{\psi_{2}+\frac{\pi}{2}})\big].
\end{eqnarray}
Now, by using that
\begin{eqnarray}
\sigma_{\psi_{2}+\frac{\pi}{2}}=\cos\psi_{21}\sigma_{\psi_{1}+\frac{\pi}{2}}
+\sin\psi_{21}\sigma_{\psi_{1}+{\pi}},
\end{eqnarray}
where $\psi_{21}=\psi_2-\psi_1$, we find
\begin{eqnarray}
U_{12}=\ket{e}\bra{e}+\cos{y}(\ket{0}\bra{0}+\ket{1}\bra{1})
+i\sin{y}\sigma_{\psi} .
\end{eqnarray}
where parameter
\begin{eqnarray}
y=(\theta_{11}^2+\theta_{22}^2-2\theta_{11}\theta_{22}\cos\psi_{21})^{1/2}
\end{eqnarray}
and  operator $\sigma_{\psi}$ is a linear combination of $\sigma_x$ and $\sigma_y$.

We next simplify the remaining operator $U_{ee} \equiv \exp[-i\delta_2\pi(e^{i\varphi_2}
\ket{b_2^{\prime}}\bra{e} +\textrm{H.c.})]
\exp[-i\delta_1\pi(e^{i\varphi_1}\ket{b_1^{\prime}}\bra{e}+\textrm{H.c.})]$.
By using the same technique as above, we find
\begin{eqnarray}
U_{ee} & = & \ket{d_{12}^{\prime}}\bra{d_{12}^{\prime}} + \cos (z\pi) (\ket{e}\bra{e} +
\ket{b_{12}^{\prime}}  \bra{b_{12}^{\prime}})
\nonumber \\
 & & -i\sin (z\pi) (\ket{e}\bra{b_{12}^{\prime}} + \ket{b_{12}^{\prime}} \bra{e}),
\end{eqnarray}
with
\begin{eqnarray}
z & = & [\delta_1^2+\delta_2^2+2\delta_1\delta_2\cos(\eta^{\prime}/2)\cos{\phi_b}]^{1/2}
\nonumber \\
\ket{b_{12}^{\prime}} & = & [(1+\cos(\eta^{\prime}/2){e^{i\phi_b}}) e^{i\varphi_{1}}\ket{b_1^{\prime}}
\nonumber \\
 & &
+\sin(\eta^{\prime}/2){e^{i\phi_d}}\ket{d_1^{\prime}}]/z .
\end{eqnarray}
Here, $\eta^{\prime}$ is the angle between the two state vectors $\ket{b_1^{\prime}}$ and
$\ket{b_2^{\prime}}$, and $\phi_b$ and $\phi_d$ are phases.

By combining $U_{12}$ and $U_{ee}$, we can now simplify the fidelity $F^{\prime\prime}$.
Since both $|y| \ll 1$ and $|z| \ll 1$, we ignore higher-order terms and the fidelity
reads
\begin{eqnarray}
F^{\prime\prime}=1-\frac{1}{3}y^2-\frac{1}{3}\pi^2z^2.
\end{eqnarray}
As a consistency check, one may note that $F^{\prime\prime}$ turns into $F^{\prime}$
when $\epsilon_0 = \epsilon_1$, i.e., the Rabi frequency errors of
the two pulses coincide. Specifically, the term $y^2/3$ vanishes and the term $\pi^2z^2/3$
turns into $f_1\pi^2\epsilon^2/3$ when $\epsilon_0=\epsilon_1 \equiv \epsilon$.

We now analyze how to find the path that optimizes the robustness the fidelity $F^{\prime\prime}$.
To this end, we rewrite $\epsilon_0$ and $\epsilon_1$ as
\begin{eqnarray}
\epsilon_0=\epsilon+\kappa, \ \ \ \epsilon_1=\epsilon-\kappa ,
\end{eqnarray}
where $\epsilon$ and $\kappa$ are the average error and relative error difference, respectively.
In the case where $\kappa = 0$, we have already shown that sensitivity of the fidelity to
$\epsilon$ can be minimized by choosing a path characterized by the phase $\phi_b=\pi$.
For nonvanishing relative error difference, we need to find a path that minimizes the $\kappa$
dependence as well. This amounts to solving
\begin{eqnarray}
\left. \frac{\partial{F^{\prime\prime}(\epsilon, \kappa)}}{\partial{\kappa}}
\right|_{\kappa=0} & = & -\frac{2}{3}(1-\cos\frac{\eta}{2})(\cos\theta_1 +
\cos \theta_2)\pi^2 \epsilon
\nonumber \\
 & = & 0 ,
\end{eqnarray}
for all $\eta$. Thus, the optimization condition reads
\begin{eqnarray}
\cos\theta_1+\cos\theta_2=0.
\end{eqnarray}
As we know, $\theta_1, \theta_2 \in [0, \pi]$. Thus the meaning of the above condition is
\begin{eqnarray}
\theta_1=\frac{\pi}{2}+\varsigma, \ \ \ \theta_2=\frac{\pi}{2}-\varsigma,
\end{eqnarray}
where $\varsigma$ is an angle depending on the realized gate.

To sum up, in the case of nonvanishing relative error difference $\kappa \neq 0$, the
robustness of the nonadiabatic holonomic gate is optimized by a path characterized by
\begin{eqnarray}
\phi_b=\pi, \ \ \ \cos\theta_1+\cos\theta_2 = 0.
\end{eqnarray}
Here, $\phi_b$ is a phase, and $\theta_1$ and $\theta_2$ are ratio angles of the Rabi
frequencies.

\section{Conclusions}
Different evolution paths can be used to implement a given nonadiabatic holonomic gate.
These paths are equivalent in the ideal case, but behave differently in the presence of errors.
Aiming to optimize the robustness with respect to the path, we have examined a selection
of evolution paths in the presence of Rabi frequency errors, which is a major error source 
in laser-driven implementations of quantum gates. These belong to three different kinds of 
realizations of nonadiabatic holonomic one-qubit gates in the three-level $\Lambda$ 
configuration: the original two-loop scheme \cite{Sjoqvist}, the single-loop multiple-pulse 
scheme \cite{Herterich}, and the single-shot scheme \cite{Xu3,Sjoqvist2}, in the presence 
of systematic errors in the Rabi frequencies.

We have found that the optimal path belongs to the two-loop scheme in the presence of
systematic errors in the Rabi frequencies. More importantly, we have also
shown that the robustness of the gates can be optimized by certain parameter choices
in the two-loop scheme. Our investigation applies to cases where the evolution period 
is shorter than the coher- ence time, and may help to strengthening the robustness of 
nonadiabatic holonomic quantum computation with respect to evolution path.

\section*{Acknowledgments}
G.F.X. acknowledges support from the National Natural Science Foundation of China
through Grants No. 11547245 and No. 11605104, from the Future Project for
Young Scholars of Shandong University through Grant No. 2016WLJH21, and from the Carl Tryggers Stiftelse (CTS) through Grant No. 14:441. P.Z.Z. acknowledges
support from the National Natural Science Foundation of China through Grant No. 11575101.
D.M.T. acknowledges support from the National Basic Research Program of China through
Grant No. 2015CB921004. E.S. acknowledges financial support from the Swedish Research
Council (VR) through Grant No. D0413201.

\end{document}